# Highly spin-polarized conducting state
# at the interface between non-magnetic band insulators: LaAlO$_3$/FeS$_2$ (001)

J. D. Burton[*] and E. Y. Tsymbal[**]

*Department of Physics and Astronomy, Nebraska Center for Materials and Nanoscience, University of Nebraska, Lincoln, Nebraska 68588-0299, USA*

First-principles density functional calculations demonstrate that a spin-polarized two-dimensional conducting state can be realized at the interface between two non-magnetic band insulators. The (001) surface of the diamagnetic insulator FeS$_2$ (pyrite) supports a localized surface state deriving from Fe $d$-orbitals near the conduction band minimum. The deposition of a few unit cells of the polar perovskite oxide LaAlO$_3$ leads to electron transfer into these surface bands, thereby creating a conducting interface. The occupation of these narrow bands leads to an exchange splitting between the spin sub-bands, yielding a highly spin-polarized conducting state distinct from the rest of the non-magnetic, insulating bulk. Such an interface presents intriguing possibilities for spintronics applications.

With the ever approaching scaling and power consumption limit of current semiconductor device technology, the search is on for new materials systems which could form the basis of the next of generation technology.[1] Going beyond traditional semiconductors to more exotic materials, such as complex oxides[2] and transition metal sulfides[3,4], could lead to lower power consumption and better scalability by offering more functionality based on various magnetic and electric degrees of freedom.[5] This is especially true for atomically engineered interfaces where new properties can be found that even the bulk constituents do not possess.[6]

One of the most prominent systems is the two-dimensional electron gas (2DEG) formed at the (001) interface between the two insulating perovskite oxides, LaAlO$_3$ and SrTiO$_3$.[7] Due to this interface being polar, charge is transferred to the interface to eliminate the internal electric field, leading to a 2DEG above a certain critical thickness of LaAlO$_3$.[8] Tunable metallic properties of this interface have spurred much interest, promising the potential for applications.[9-13] In addition, magnetism[14,15] and superconductivity[14,15] have been discovered at these interfaces, suggesting further implications for nanoelectronics.[6]

Making a spin-polarized 2DEG is an exciting prospect for spintronics, where the involvement of the spin degree of freedom broadens the spectrum of potential applications.[16] In order to incorporate magnetism at conducting interfaces several systems have been proposed and studied, e.g. replacing LaAlO$_3$ with the strongly correlated oxide LaVO$_3$,[17] imbedding a LaO monolayer in SrMnO$_3$,[18] and exploiting ferromagnetism of EuO.[19,20] All of these interfaces inherit their magnetic properties from the constituent bulk materials, either through their magnetic order or their tendency toward strong electron correlations.

Here we propose a different approach to create a spin-polarized 2DEG: magnetism is induced at the interface of two non-magnetic insulators due to the exchange splitting of the interface states driven by charge transfer to this interface. Such an interface can be realized by pairing a polar LaAlO$_3$ with the diamagnetic band insulator iron disulfide, FeS$_2$, commonly known as pyrite. FeS$_2$ begins a series of pyrite-structure disulfides covering every member of the late half of the 3$d$ elements all the way to ZnS$_2$, each displaying unique properties distinct from their neighbors.[21] In particular, CoS$_2$ has one more $d$ electron per formula unit compared to FeS$_2$ which makes it an itinerant ferromagnetic metal. Changing this charge through alloying of CoS$_2$ and FeS$_2$ allows tuning spin polarization and other magnetic and transport properties.[3,22] This suggests that by, alternatively, electron doping of a pure FeS$_2$ surface through heterostructuring with polar LaAlO$_3$ one might achieve both conductivity and magnetism at the same interface.

To this end we present here results of first-principles density functional theory (DFT) calculations of LaAlO$_3$/FeS$_2$ (001) interfaces that confirm the conducting and ferromagnetic behaviors at such an interface. These properties are confined to the interface due to the presence of native surface states of FeS$_2$. These states are highly susceptible to the Stoner exchange splitting when occupied, giving rise to itinerant ferromagnetism and a substantial spin polarization.

DFT calculations of atomic and electronic structure are performed using the plane-wave pseudopotential method implemented in the Quantum ESPRESSO package.[23] A plane-wave cutoff energy of 400 eV and a generalized-gradient approximation (GGA)[24] were used in all the calculations. Atomic relaxations were converged using a 4×4×1 Monkhorst-Pack $k$-point mesh, a Gaussian broadening of 0.1 eV and a force cutoff of 20 meV/Å. The resulting structures were used in subsequent frozen-lattice self-consistent calculations using a denser 10×10×1 $k$-point mesh and a broadening of 0.02 eV to further refine the electronic charge density. Subsequent non-self-consistent calculations on a 48×48×1 $k$-point mesh were performed to extract $k_\parallel$-resolved local density of states (LDOS) with 7 meV broadening.

Bulk $FeS_2$ (pyrite) has a quasi-rocksalt cubic structure consisting of $Fe^{2+}$ at the face centers and $S_2^{2-}$ dimers centered at the cube corners and alternately aligned along the various equivalent body-diagonal axes, resulting in space group $Pa\bar{3}$. Our calculations yield a cubic lattice constant of $a = 5.410$ Å and a $S_2^{2-}$ bond length of $d = 2.194$ Å, in good agreement with experimentally measured values of $a = 5.416$ Å and $d = 2.12$ Å.[25]

We study three related heterostructure systems, shown in Fig. 1, using a tetragonal supercell. In all cases the in-plane lattice parameter is fixed to the calculated (GGA) lattice constant of bulk $FeS_2$ to mimic epitaxial growth on a single crystal or well relaxed film. The vertical supercell size is $13a = 70.33$ Å. First we study the bare slab consisting of 5 stoichiometric (001) layers of $FeS_2$ embedded in vacuum as shown in Fig. 1a. Other surface terminations are energetically unfavorable, making (001) an ideal cleavage plane for single crystals resulting in flat, atomically stepped, terraces up to a few hundred nm wide.[26] Relaxing atomic coordinates in this slab does not introduce dramatic changes to the structure with respect to the bulk, consistent with previous calculations[27,28] and experimental data.[26,29]

In bulk, $Fe^{2+}$ cations are 6-fold coordinated by sulfur. The crystal field is octahedral, splitting the 3$d$ manifold into a low-lying $t_{2g}$ triplet and a higher energy $e_g$ doublet. This splitting is large enough that the zero-spin state is favored with 6 electrons accommodated in the $t_{2g}$ orbitals, leaving the $e_g$ orbitals above the band gap. This behavior is evident from the LDOS of the bulk-like $FeS_2$-3 layer in Fig. 2a. On the (001) surface $Fe^{2+}$ cations are only 5-fold coordinated, modifying the crystal field environment of the Fe-3$d$ states. The $e_g$ doublet is split and the $t_{2g}$ states are split into a low singlet and a higher doublet. These split levels alone do not close the gap, leaving 6 spin-paired electrons in the split "$t_{2g}$" sector of the manifold. Since this change in crystal splitting is localized at the surface, however, the high-lying levels of the split "$t_{2g}$" triplet and the low-lying member of the split "$e_g$" doublet constitute surface states near the top of the valence band and bottom of the conduction band, respectively.

The signature of these surface states is seen in Fig. 2a as peaks at around $E_F \pm 0.4$ in the LDOS which are strongest on the surface layer, $FeS_2$-1, but quickly decay into the sub-surface layers. The decay of the Fe conduction band surface states can also be seen in the $k_\parallel$- and layer-resolved LDOS plotted in Figs. 3a-c at $E_F + 0.4$ eV. The narrow energy contours correspond to cuts through a two-dimensional band structure, demonstrating a decrease in intensity when moving from the surface (Fig.3a) to the bulk (Fig.3c).

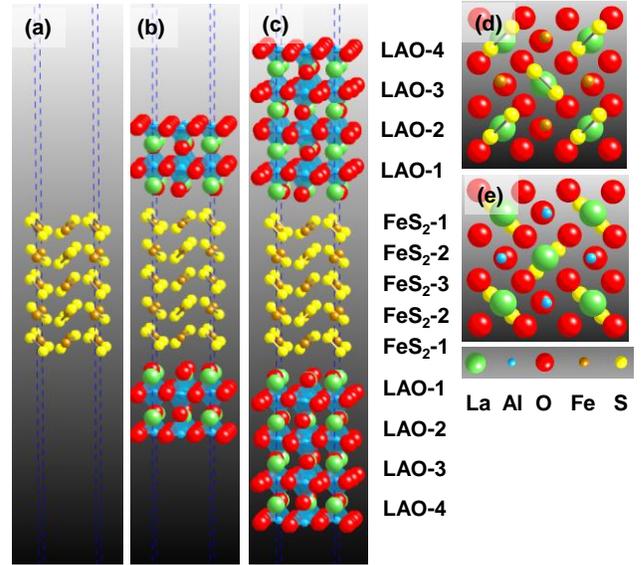

FIG 1. Atomic structures of the three systems studied: (a) $FeS_2$ slab consisting of five (001) atomic layers; (b,c) symmetric $LaAlO_3/FeS_2$ heterostructures with the $FeS_2$ slab covered by 2 (b) and 4 (c) u.c. $LaAlO_3$ films; (d,e) bottom (d) and top (e) views of the first few monolayers of the interface structures.

Next, we study the $LaAlO_3/FeS_2$ (001) interface. Bulk $LaAlO_3$ deviates from the perfect cubic perovskite structure by the presence of tilts and rotations of the oxygen octahedral cages around the Al sites, resulting in a rhombohedral structure with space group $R\bar{3}c$. In epitaxial films, grown along the [001] pseudocubic direction on a cubic substrate, biaxial strain induces a change in symmetry dependent on the sign and magnitude of the strain.[30] Our GGA calculations of the bulk (unconstrained) $R\bar{3}c$ structure reveal a volume consistent with a cubic perovskite lattice parameter $a_{cp} = 3.817$ Å. Epitaxial matching with the pyrite structure requires a $\sqrt{2}\times\sqrt{2}$ in-plane doubling of the pseudocubic perovskite cell with a 45° rotation around the pseudocubic [001] direction, leading to an effective in-plane lattice constant of $\sqrt{2}a_{cp} = 5.398$ Å. Matching to the $FeS_2$ lattice leads to a small -0.2% tensile strain, and GGA calculations of bulk $LaAlO_3$ for this strain state reveal a $C2/c$ structure, qualitatively consistent with the previous calculations.[30]

Using this $C2/c$ structure as a starting point we construct the heterostructures by adding 2 unit cell (u.c.) and 4 u.c. $LaAlO_3$ layers to the $FeS_2$ slab, as shown in Figs. 1b and 1c, respectively. In both cases the $LaAlO_3$ films are stoichiometric with LaO termination at the interfaces with $FeS_2$ and $AlO_2$ termination with vacuum. The LaO interface termination, with $La^{3+}$ just above the center of the $S_2^{2-}$ dimers and $O^{2-}$ just above the $Fe^{2+}$ sites, is a natural extension of the rock-salt-like ionic structure of the $FeS_2$ surface (Figs. 1d-e). Each supercell is inversion symmetric, eliminating any macroscopic electric fields in the vacuum regions. Supercells are taken sufficiently large to minimize any interactions across the vacuum. The structures are then fully relaxed.

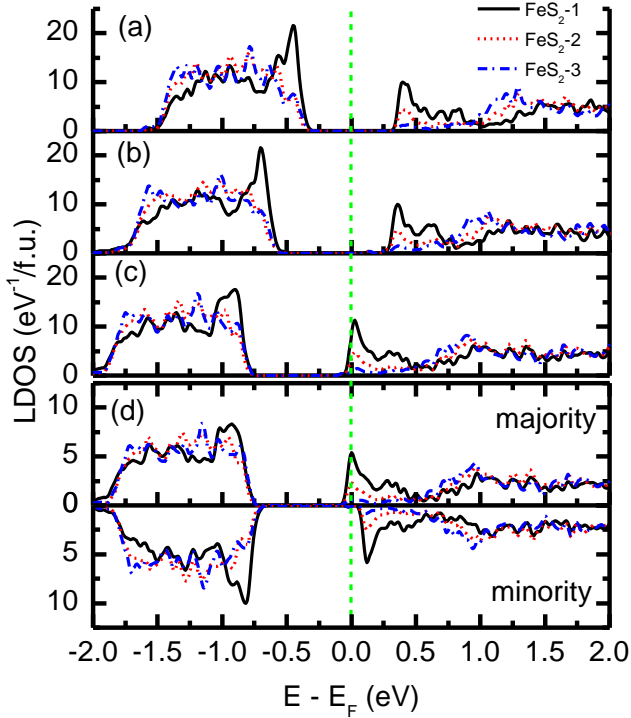

FIG 2. LDOS projected onto layers $FeS_2$-1 through -3 (as denoted in Fig. 1) for a $FeS_2$ slab surrounded by (a) vacuum, (b) 2 u.c. $LaAlO_3$ and (c-d) 4 u.c. $LaAlO_3$, as follows from non-spin polarized (c) and spin-polarized (d) calculations. The vertical dashed line indicates the Fermi energy.

We find that the 2 u.c. $LaAlO_3$ system maintains a true band gap and thus remains insulating. Nevertheless it is evident from Fig. 2b that the electric field in the $LaAlO_3$ has shifted the conduction band minimum, and the surface states in $FeS_2$ are closer to the Fermi level than those for the bare $FeS_2$ slab (Fig. 2a). This tendency persists with increasing $LaAlO_3$ thickness, and for the 4 u.c. $LaAlO_3$ heterostructure we find the Fermi level lying within the conduction band of $FeS_2$, thus indicating metallicity of the interface (see Fig. 2c). This behavior is consistent with the charge transfer to the interface above a $LaAlO_3$ critical thickness known for the well-studied $LaAlO_3/SrTiO_3$ system.[31-33]

A crucial difference of the $LaAlO_3/FeS_2$ interface is, however, the fact that the transferred electrons are almost entirely accommodated into the well-localized Fe-$e_g$ surface states of $FeS_2$. This is seen from the $k_\parallel$-resolved LDOS in $FeS_2$ plotted in Figs. 3d-f, where the narrow contours correspond to the Fermi surface sheets of this two-dimensional conducting interface, which are very similar to the $FeS_2$ bare surface state in Figs. 3a-c.

The above calculation assumed no spin polarization.

This constraint results in a relatively large peak in the non-spin polarized LDOS at the Fermi level on the surface Fe atoms (Fig. 3c). This suggests the possibility of exchange splitting of the spin bands to reduce electron energy.[34] Our spin-polarized calculation confirms this prediction. Fig. 4 shows the planar averaged spin density profile revealing that the dominant contribution to magnetic moment comes from the Fe sites in the $FeS_2$-1 layer, whereas the magnetization in the rest of the structure is nearly negligible. The corresponding induced magnetization is 0.13 $\mu_B$ per interface Fe.

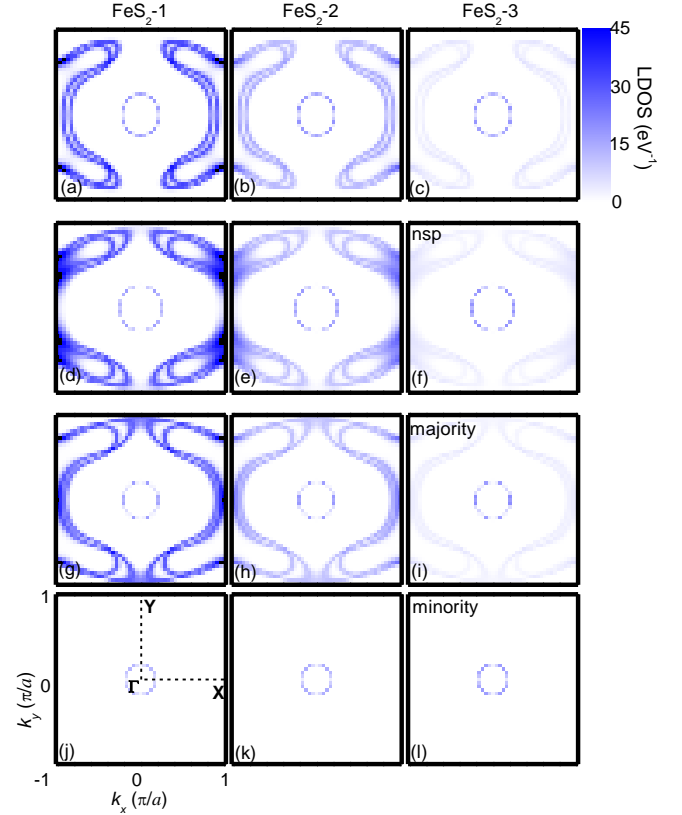

FIG 3. $k_\parallel$-resolved LDOS projected onto layers $FeS_2$-1 through -3 (as denoted in Fig. 1) for a $FeS_2$ slab, $E = E_F + 0.4$ eV (a-c) and $LaAlO_3$(4 u.c.)/$FeS_2$ heterostructure, $E = E_F$ (d-l). Results of non-spin-polarized (d-f) and spin-polarized calculations for majority (g-i) and minority spin (j-l) are shown.

The calculated exchange splitting of the surface states is 0.11 eV. As seen from Fig. 2d, this completely splits the Fe-$d$ states making the system nearly half-metallic with Fermi-level LDOS dominated by the majority-spin states. The spin- and $k_\parallel$-resolved LDOS in $FeS_2$ are plotted in Figs. 3g-l. The majority-spin LDOS (Figs. 3g-i) looks similar to those for the non-spin polarized interface states and the bare surface states of $FeS_2$. The minority-spin LDOS (Figs. 3j-l) displays a small 2-dimensional electron pocket which decays slowly into the bulk.

We use the Stoner model for itinerant ferromag-netism [34] to explain the calculated magnetic moment and exchange splitting. According to this model exchange splitting Δ of the spin bands forms a spin magnetic moment $m = \int_{\varepsilon_F - \Delta_2}^{\varepsilon_F + \Delta_1} \mu_B \rho(\varepsilon) d\varepsilon$. Here $\varepsilon_F$ is the paramagnetic Fermi energy, ρ(ε) is the LDOS per spin in the paramagnetic state, and $\Delta_1$ and $\Delta_2$ denote the exchange driven shifts of the majority- and minority-spin bands so that $\Delta_1 + \Delta_2 = \Delta$. The total energy $U$ is the sum of the band energy, $U_b$, and the exchange energy, $U_{ex}$, i.e. $U = U_b + U_{ex} = \int_{\varepsilon_F - \Delta_2}^{\varepsilon_F + \Delta_1} \varepsilon \rho(\varepsilon) d\varepsilon - Im^2/4\mu_B^2$, where $I$ is the Stoner exchange parameter. These two competing contributions have to be balanced in order to determine the equilibrium magnetic moment, resulting in the well-known relationship $\Delta = Im$.[34-36] Taking $m = 0.13$ $\mu_B$ and $\Delta = 0.11$ eV from the DFT calculations we find an exchange parameter $I = 0.84$ eV.

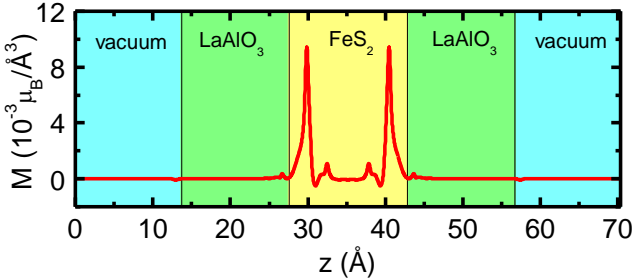

FIG 4. Distribution of spin-magnetization, $M$, in the 4 unit-cell case averaged over the plane parallel to the layers.

Next, we performed a series of Stoner model calculations based on the paramagnetic LDOS on the $FeS_2$ surface layer, ρ(ε). By minimizing the total energy $U$ with appropriate constraints on charge conservation we determined the equilibrium moments as a function of $I$, as plotted in Fig. 5a. We found an abrupt turn on of magnetization around $I = 0.47$ eV corresponding to satisfaction of the Stoner criterion for magnetism, $I\rho(\varepsilon_F) > 1$, where in our case $\rho(\varepsilon_F) = 2.1$ eV$^{-1}$. We also found that the Fe magnetic moment quickly saturates to $m \sim 0.10$ $\mu_B$ corresponding to the half-metallic state where charge in the minority-spin channel is completely depleted. This value is less than $m = 0.13$ $\mu_B$ found from the DFT calculation due to our assumption of all magnetization and band energy originating solely from the $FeS_2$ surface states. Next, we performed a series of constrained-moment DFT calculations to determine the value of $I$. We computed the total energy, $U$, versus small, fixed magnetic moment $m$, plotted in Fig. 5b. Then we determined the Stoner exchange energy, $U_{ex}$, as a function of $m$ by subtracting from this curve the band energy, $U_b$, as determined by shifting the paramagnetic LDOS, ρ(ε). Fitting $U_{ex}$ with a parabola we obtained the Stoner exchange parameter $I = 0.81$ eV. This value is remarkably close to the value $I = 0.84$ eV we found above from the Stoner equilibrium criterion indicating that our classification of the interface magnetism as *itinerant* is appropriate.

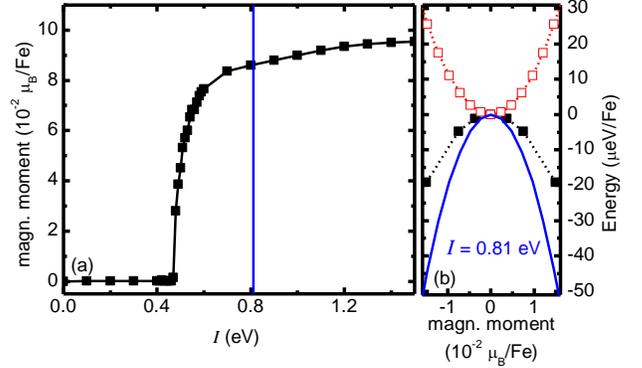

FIG 5. (a) Equilibrium magnetic moments from the Stoner model calculation as a function of exchange parameter $I$. Symbols connected by lines are results of numerical calculations and the vertical line indicates the calculated value of $I$ from DFT data in (b). (b) $U$ vs. $m$ data used to determine the exchange parameter $I$. The solid symbols are total energies from spin-polarized DFT calculations with fixed total magnetization. Open symbols are calculated $U_b$. Dotted curves are a guide to the eyes. The solid curve is $U_{ex}$: the difference between parabolic fits to the $U$ and the $U_b$.

In conclusion we have predicted a conducting ferromagnetic interface between two non-magnetic band insulators, $LaAlO_3$ and $FeS_2$. The formation of this interface is driven by the polar nature of the $LaAlO_3$ (001) layer that supports charge transfer to a localized surface state formed by Fe $d$-orbitals at the conduction band minimum of $FeS_2$. This nearly half-metallic interface may be interesting for spintronics applications.

This work was supported by the Nebraska MRSEC (NSF Grant No. DMR-0820521) and NSF-EPSCoR (Grant No. EPS-1010674). Computations were performed utilizing the Holland Computing Center of the University of Nebraska.

[*]e-mail: jdburton1@gmail.com [**]e-mail: tsymbal@unl.edu

[1]International Technology Roadmap for Semiconductors. (2003). Semicond. Ind. Assoc., San San Jose, CA. [Online]. Available: http://public.itrs.net
[2]Y. Tokura and H. Y. Hwang, Nat Mater **7**, 694 (2008).
[3]I. I. Mazin, Appl. Phys. Lett. **77**, 3000 (2000).
[4]L. Wang, K. Umemoto, R. M. Wentzcovitch, T. Y. Chen, C. L. Chien, J. G. Checkelsky, J. C. Eckert, E. D. Dahlberg, and C. Leighton, Phys. Rev. Lett. **94**, 056602 (2005).
[5]J. P. Velev, P. A. Dowben, E. Y. Tsymbal, S. J. Jenkins, and A. N. Caruso, Surface Science Reports **63**, 400 (2008).
[6]J. Mannhart and D. G. Schlom, Science **327**, 1607 (2010).
[7]A. Ohtomo and H. Y. Hwang, Nature **427**, 423 (2004).
[8]N. Nakagawa, H. Y. Hwang, and D. A. Muller, Nature Mater. **5**, 204 (2006).
[9]S. Thiel, G. Hammerl, A. Schmehl, C. W. Schneider, and J. Mannhart, Science **313**, 1942 (2006).
[10]C. Cen, S. Thiel, J. Mannhart, and J. Levy, Science **323**, 1026 (2009).

[11] M. K. Niranjan, Y. Wang, S. S. Jaswal, and E. Y. Tsymbal, Phys. Rev. Lett. **103**, 016804 (2009).

[12] C. W. Bark, D. A. Felker, Y. Wang, Y. Zhang, H. W. Jang, C. M. Folkman, J. W. Park, S. H. Baek, H. Zhou, D. D. Fong, X. Q. Pan, E. Y. Tsymbal, M. S. Rzchowski, and C. B. Eom, Proceedings of the National Academy of Sciences **108**, 4720 (2011).

[13] H. W. Jang, D. A. Felker, C. W. Bark, Y. Wang, M. K. Niranjan, C. T. Nelson, Y. Zhang, D. Su, C. M. Folkman, S. H. Baek, S. Lee, K. Janicka, Y. Zhu, X. Q. Pan, D. D. Fong, E. Y. Tsymbal, M. S. Rzchowski, and C. B. Eom, Science **331**, 886 (2011).

[14] A. Brinkman, M. Huijben, M. van Zalk, J. Huijben, U. Zeitler, J. C. Maan, W. G. van der Wiel, G. Rijnders, D. H. A. Blank, and H. Hilgenkamp, Nature Mater. **6**, 493 (2007).

[15] N. Reyren, S. Thiel, A. D. Caviglia, L. F. Kourkoutis, G. Hammerl, C. Richter, C. W. Schneider, T. Kopp, A. S. Ruetschi, D. Jaccard, M. Gabay, D. A. Muller, J. M. Triscone, and J. Mannhart, Science **317**, 1196 (2007).

[16] I. Zutic, J. Fabian, and S. D. Sarma, Rev. Mod. Phys. **76**, 323 (2004).

[17] Y. Hotta, T. Susaki, and H. Y. Hwang, Phys. Rev. Lett. **99**, 236805 (2007).

[18] B. R. K. Nanda and S. Satpathy, Phys. Rev. Lett. **101**, 127201 (2008).

[19] Y. Wang, M. K. Niranjan, J. D. Burton, J. M. An, K. D. Belashchenko, and E. Y. Tsymbal, Phys. Rev. B **79**, 212408 (2009).

[20] J. Lee, N. Sai, and A. A. Demkov, Phys. Rev. B **82**, 235305 (2010).

[21] H. S. Jarrett, W. H. Cloud, R. J. Bouchard, S. R. Butler, C. G. Frederick, and J. L. Gillson, Phys. Rev. Lett. **21**, 617 (1968).

[22] L. Wang, T. Y. Chen, C. L. Chien, J. G. Checkelsky, J. C. Eckert, E. D. Dahlberg, K. Umemoto, R. M. Wentzcovitch, and C. Leighton, Phys. Rev. B **73**, 144402 (2006).

[23] P. Giannozzi, S. Baroni, N. Bonini, M. Calandra, R. Car, C. Cavazzoni, D. Ceresoli, G. L. Chiarotti, M. Cococcioni, I. Dabo, A. D. Corso, S. d. Gironcoli, S. Fabris, G. Fratesi, R. Gebauer, U. Gerstmann, C. Gougoussis, A. Kokalj, M. Lazzeri, L. Martin-Samos, N. Marzari, F. Mauri, R. Mazzarello, S. Paolini, A. Pasquarello, L. Paulatto, C. Sbraccia, S. Scandolo, G. Sclauzero, A. P. Seitsonen, A. Smogunov, P. Umari, and R. M. Wentzcovitch, J. Phys.: Cond. Mat. **21**, 395502 (2009).

[24] J. P. Perdew, K. Burke, and M. Ernzerhof, Phys. Rev. Lett. **77**, 3865 (1996).

[25] S. L. Finklea, III, L. Cathey, and E. L. Amma, Acta. Crystallogr. Sect. A **32**, 529 (1976).

[26] K. M. Rosso, U. Becker, and M. F. Hochella, Am. Miner. **84**, 1535 (1999).

[27] A. Hung, J. Muscat, I. Yarovsky, and S. P. Russo, Surf. Sci. **513**, 511 (2002).

[28] A. Stirling, M. Bernasconi, and M. Parrinello, J. Chem. Phys. **118**, 8917 (2003).

[29] K. M. Rosso, Rev. Miner. Geochem. **42**, 199 (2001).

[30] A. J. Hatt and N. A. Spaldin, Phys. Rev. B **82**, 195402 (2010).

[31] J. Lee and A. A. Demkov, Phys. Rev. B **78**, 193104 (2008).

[32] R. Pentcheva and W. E. Pickett, Phys. Rev. Lett. **102**, 107602 (2009).

[33] H. Chen, A. Kolpak, and S. Ismail-Beigi, Phys. Rev. B **82**, 085430 (2010).

[34] E. C. Stoner, Proc. R. Soc. A **165**, 372 (1938).

[35] P. M. Marcus and V. L. Moruzzi, Phys. Rev. B **38**, 6949 (1988).

[36] Y. Sun, J. D. Burton, and E. Y. Tsymbal, Phys. Rev. B **81**, 064413 (2010).